\documentclass{emulateapj}

\def\mathstacksym#1#2#3#4#5{\def#1{\mathrel{\hbox to 0pt{\lower#5\hbox{#3}\hss} \raise #4\hbox{#2}}}}
\mathstacksym\gta{$>$}{$\sim$}{1.5pt}{3.5pt} 
\mathstacksym\lta{$<$}{$\sim$}{1.5pt}{3.5pt} 
\begin{document}
                                %
\title{VLTI/MIDI 10 micron interferometry of the forming massive star W33A\thanks{Based on
    observations with VLTI, proposal 075.C-0703}}
\shorttitle{MIDI observations of W33A}
\author{W.J. de Wit\altaffilmark{1}, M.G. Hoare\altaffilmark{1},
  R.D. Oudmaijer\altaffilmark{1}, J.C. Mottram\altaffilmark{1}} 
\altaffiltext{1}{School of Physics and Astronomy, University of Leeds, LS2 9JT, UK; phywjmdw@ast.leeds.ac.uk, mgh@ast.leeds.ac.uk, roud@ast.leeds.ac.uk, jcm@ast.leeds.ac.uk}          
                            
\begin{abstract}

We report on resolved interferometric observations with VLTI/MIDI of
the massive young stellar object (MYSO) W33A. The MIDI observations
deliver spectrally dispersed visibilities with values between 0.03 and
0.06, for a baseline of 45m over the wavelength range 8-13$\mu$m. The
visibilities indicate that W33A has a FWHM size of approximately
120\,AU (0.030'') at 8$\mu$m which increases to 240\,AU at 13$\mu$m,
scales previously unexplored among MYSOs. This observed trend is
consistent with the temperature falling off with distance. 1D dust
radiative transfer models are simultaneously fit to the visibility
spectrum, the strong silicate feature and the shape of the mid
infrared spectral energy distribution (SED). For any powerlaw density
distribution, we find that the sizes (as implied by the visibilities)
and the  stellar luminosity are incompatible. A reduction to a
third of W33A's previously adopted luminosity is required to match the
visibilities; such a reduction is consistent with new high resolution
70$\mu$m data from Spitzer's MIPSGAL survey. We obtain best fits for
models with shallow dust density distributions of $r^{-0.5}$ and
$r^{-1.0}$ and for increased optical depth in the silicate feature
produced by decreasing the ISM ratio of graphite to silicates and
using optical grain properties by Ossenkopf et al. (1992).
\end{abstract}


\section{Introduction}
Massive young stellar objects are luminous, embedded infrared (IR)
sources that show many signs that they are still actively accreting
mass. Their luminosity ($ L>10^{4}L_{\odot}$) is such that they are
expected to be ionizing their surroundings to produce an H II region,
yet they only have weak radio emission due to ionized winds or jets
(Hoare 2002\nocite{2002ASPC..267..137H}). Most appear to be driving
bipolar molecular flows and (sub-)millimetre interferometry is beginning to reveal evidence of rotating, disc-like
structures on scales of hundreds of AU (Patel et
al. 2005\nocite{2005Natur.437..109P}; Beltr\'{a}n et
al. 2006\nocite{2006Natur.443..427B}; Torrelles et
al. 2007\nocite{2007ApJ...666L..37T}). There is great interest in
knowing the distribution of the infalling and outflowing material on
smaller scales to provide clues to which physical processes are controlling
the dynamics and setting the final mass of the star (Beuther et
al. 2007\nocite{2007prpl.conf..165B}; Hoare et
al. 2007\nocite{2007prpl.conf..181H}).

The mid-IR (8-13$\mu$m) emission from MYSOs is thought to arise in the
warm ($\sim$300\,K) dust in the envelope heated by the young
star. Previous modelling (e.g. Churchwell et al. 1990\nocite{1990ApJ...354..247C}) has indicated
that the size of the mid-IR emission region should be unresolved at
the typical distances of MYSOs by single-dishes and this is borne out
by observations (Kraemer et al. 2001\nocite{2001ApJ...561..282K}; Mottram et al. 2007\nocite{2007arXiv0709.2040M}). Exceptions
occur when MYSOs are viewed close to edge-on when a dense torus can
completely obscure the bright central region and only the warm dust in
the outflow cavities is seen (de Buizer 2006\nocite{2006ApJ...642L..57D}).

Here we present new results from mid-IR interferometric observations of the MYSO
W33A using the VLTI MIDI instrument. W33A has a kinematic distance of 3.8 kpc
and a luminosity as derived from IRAS fluxes of $\rm 1 \times 10^{5}
L_{\odot}$ (Fa\'{u}ndez et al. 2004\nocite{2004A&A...426...97F}). It has weak,
compact, optically thick radio continuum emission (Rengarajan \& Ho
1996\nocite{1996ApJ...465..363R}; van der Tak \& Menten
2005\nocite{2005A&A...437..947V}) and broad ($\sim$100 km s$^{-1}$),
single-peaked H I recombination emission lines (Bunn et
al. 1995\nocite{1995MNRAS.272..346B}) consistent with an ionized stellar wind
origin. IR images from 2MASS and Spitzer's GLIMPSE survey clearly show a large
scale monopolar nebula emerging to the SE, which is most likely to demarcate
the blue-shifted lobe of a bipolar outflow. 

The cold dust envelope has been
studied at a few arcsecond resolution at millimetre wavelengths (van der Tak et
al. 2000\nocite{2000ApJ...537..283V}), probing the imprint of the star
formation process on the circumstellar dust distribution. Many different
scenarios have been put forward predicting this distribution. 
In case of W33A, G\"{u}rtler et al. (1991\nocite{}) modelled the SED
with a constant density spherical envelope and found an inner cavity
radius of 135 AU (35 mas at 3.8 kpc) with the dust close to the
sublimation temperature. MIDI provides data on comparable size scales
of 50 mas, and since the mid-IR size will be somewhat larger than the
sublimation zone we expect the emission to be well-resolved.

\section{Observations and data reduction}
\label{observ}

 \begin{figure*}
  \includegraphics[height=9cm,width=5.5cm,angle=90]{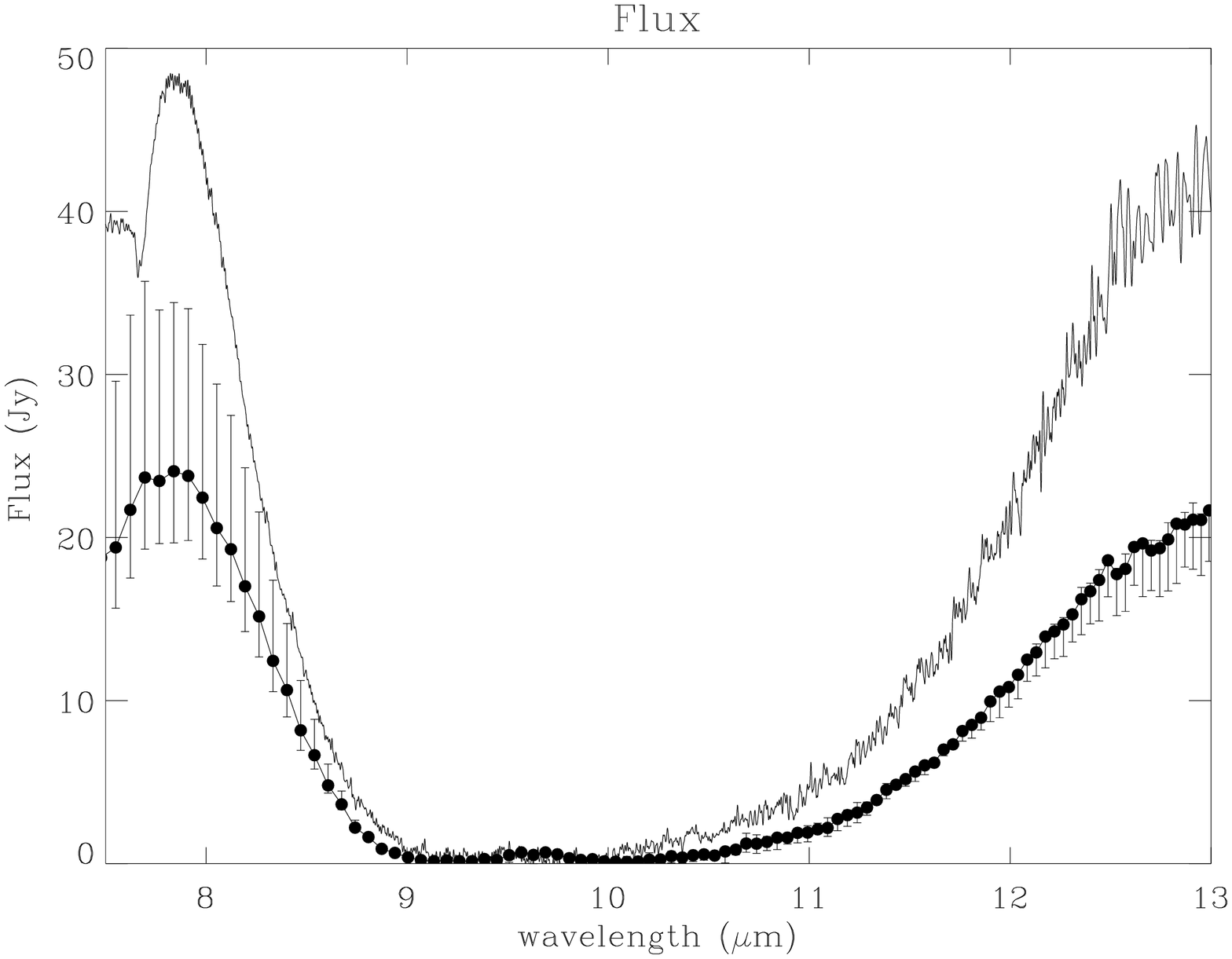}
  \includegraphics[height=9cm,width=5.5cm,angle=90]{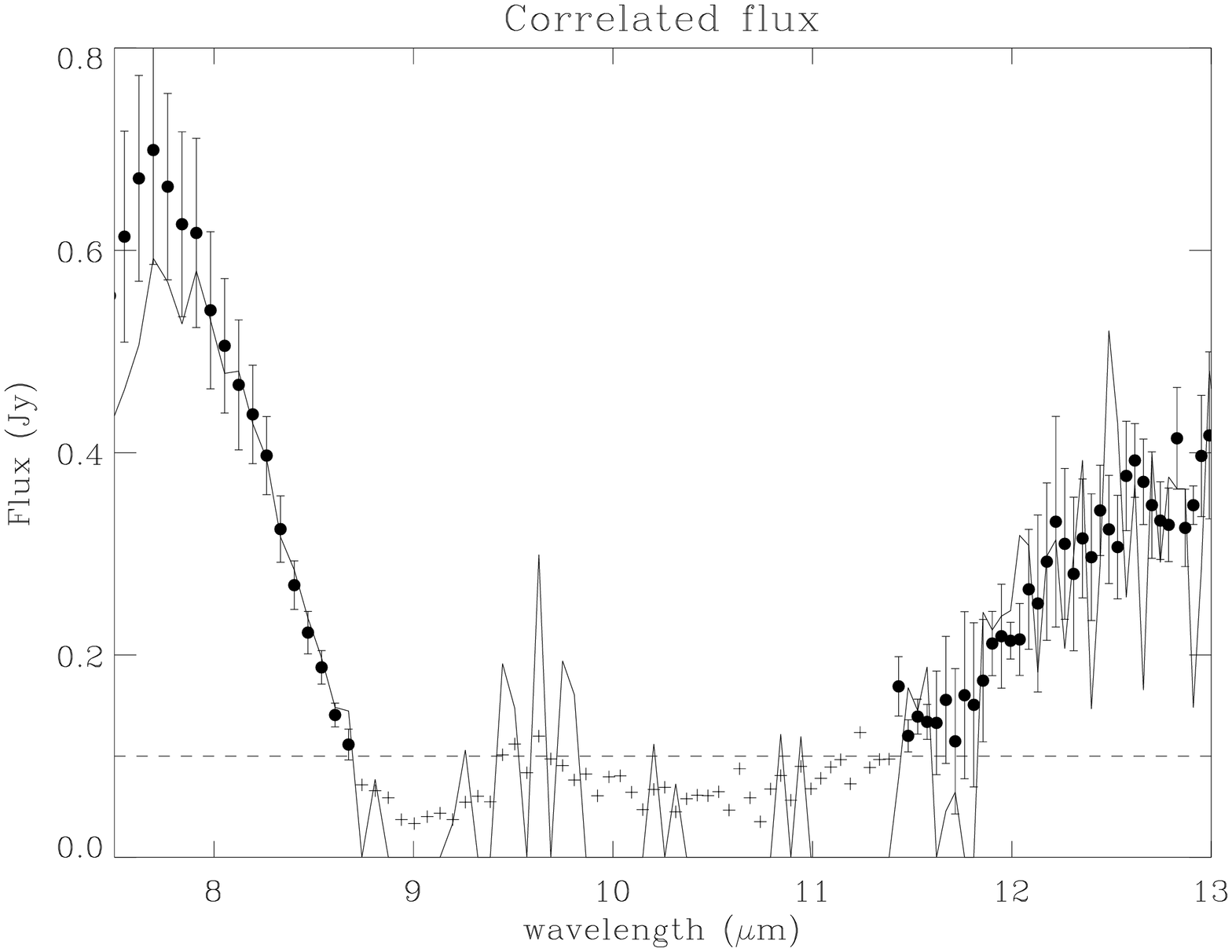}
  \caption[]{{\it Left:} MIDI (lower full line) and ISO flux calibrated
spectra of W33A.  {\it Right:} The correlated flux determined
with EWS (dots with errorbars), and MIA (full line). Note the ragged behaviour
of MIA at low flux levels. EWS errorbars correspond to the variations in
visibilities as function of reduction parameters.  EWS  and MIA are consistent
except at correlated flux levels below 0.1Jy (dashed line), these values must be
considered upper limits (see Jaffe et al. 2004\nocite{}; Matsuura et
al. 2006\nocite{2006ApJ...646L.123M}).} 
  \label{uncspec1}
\end{figure*}

MIDI is the VLTI's two telescope beam combiner that operates in the
thermal IR (Leinert et al. 2003\nocite{2003Msngr.112...13L}). W33A was
observed with MIDI on 16 September 2005, employing the UT2-UT3
telescope configuration for a projected baseline of $\rm 45.5\,m$ and
position angle of $47\degr$ east of north; this is perpendicular
to the star's outflow direction. An interferometric calibrator star
\object{HD\,169916} was observed during the same night. Lunar occultation and
indirect methods reveal the calibrator diameter to be a few mas (Nather
1972\nocite{1972PhDTnather}; Pasinetti-Fracassini
2001\nocite{2001A&A...367..521P}), too small to be resolved by MIDI. 
Here we present visibilities between 8--13$\mu$m that are spectrally dispersed at a
resolution of $\rm R=30$. The observations were executed in
the {\it High-Sens} MIDI mode (for details see Przygodda et
al. 2003\nocite{2003Ap&SS.286...85P}; Chesneau et
al. 2005\nocite{2005A&A...435..563C}; Leinert et
al. 2004\nocite{2004A&A...423..537L}).  We summarize the main
elements of the procedure that produces the raw dataset. The two beams
are interfered producing two complementary interferometric channels
that have a phase difference of $\pi$
radians.  The dispersed fringes are
modulated according to an introduced optical path delay
(OPD). Interferograms are recorded for a range in OPDs, a few
millimetre around optical path length equalization, called a fringe
scan.  For both W33A and HD\,169916 a total of 8000 interferograms,
corresponding to 200 scans were recorded. Subtracting the two interferometric beams eliminates the
background and enhances the fringe signal. The coherent flux was
extracted using two different MIDI software reduction packages: EWS (Jaffe
2004\,\nocite{2004SPIE.5491..715J}) and MIA (K\"{o}hler et
al. 2005\nocite{2005AN....326Q.563K}). MIA estimates the amplitudes in
the power spectra of the fringe signal Fourier transform (incoherent estimation). The amplitudes are proportional to the
correlated flux. EWS on the other hand aims at adding the fringes to
maximize the signal to noise (coherent estimation). This method corrects first
the fringe spectra for their corrupted phase caused by atmospheric and
instrumental effects.  EWS uses the fringe scan as
phase reference to estimate the group delay due to the
atmosphere. The observations show that the atmospheric group delay
varied within a range of 50$\mu$m over 100s, indicating relatively
good atmospheric circumstances during the night. The Fourier transform of the group
delay function reveals the typical sawtooth phase change due to the
introduced OPD, indicating that fringes have been measured. Removing
the atmospheric and the (known) instrumental group delays constitutes
a linear correction to the dispersed fringe signal, and straightens
the dispersed fringe spectra, i.e. the phase is independent of
wavelength. Next, the phase offset due to varying water refraction
index between the time of recording of the fringe spectra has to be
accounted for. In principle all spectra can then be added to a final
fringe spectrum.  Final visibilities are obtained by taking two spectra of
the source (one with each telescope) immediately after
the interferometric measurement.  The accuracy of the final visibility
spectrum is limited by the sky brightness variation between the
interferometric and flux measurement, and can amount to 10-15\%.

The flux spectra are corrected for sky contribution using a median sky
subtraction. Spectra are extracted by summing the counts within 3$\sigma$ of the mean
position, determined from a Gaussian fit to each
column along the wavelength axis. This procedure results in a W33A spectrum with an $\rm S/N
\sim 100$ and $\rm S/N \sim 300$ for the standard star. Four spectra
are recorded, 2 for each telescope beam and combined via a geometric
mean. This quantity is also what is obtained for 
the correlated flux after beam combination and thus ensures
consistency when deriving visibilities. The absolute flux calibration
is done using HD\,169916, which has an average flux density of 30Jy
  between 8$\mu$m and 13$\mu$m  (Cohen et al. 1999\nocite{}). The difference in
airmass between the observation of the flux calibrator and W33A is
0.05 leading to a negligible correction on the final fluxes. Errorbars
on the spectrum in Fig.\,\ref{uncspec1} indicate the systematic difference in flux levels
between the two telescopes beams. Absolute flux calibration is
uncertain upto at least 35\% due to difference in flux level of
HD\,169916 in each telescope beam.


\section{RESULTS}
\label{results}

Fig.\,\ref{uncspec1} presents the MIDI flux spectrum and correlated flux
spectrum from EWS and MIA. The flux spectrum is compared to the ISO-SWS
spectrum, taken from the ISO Data Archive maintained by ESA.  Both spectra are
dominated by a very strong silicate feature, that contains solid-state ammonia
and methanol absorption features, at 9 and 9.7$\mu$m respectively (see Gibb et al. 2000\nocite{2000ApJ...536..347G}). At the central wavelength
of the feature no flux was recorded, and the actual depth is unknown. The MIDI
and ISO spectra portray a similar overall shape but the MIDI spectrum has a
flux level that is about a factor 2 less. In addition to the uncertainties due
to flux calibration, we ascribe this difference to the much larger ISO beam
(80\arcsec) in comparison to MIDI (2\arcsec).

\begin{figure*}
  \includegraphics[height=9cm,width=5.5cm,angle=90]{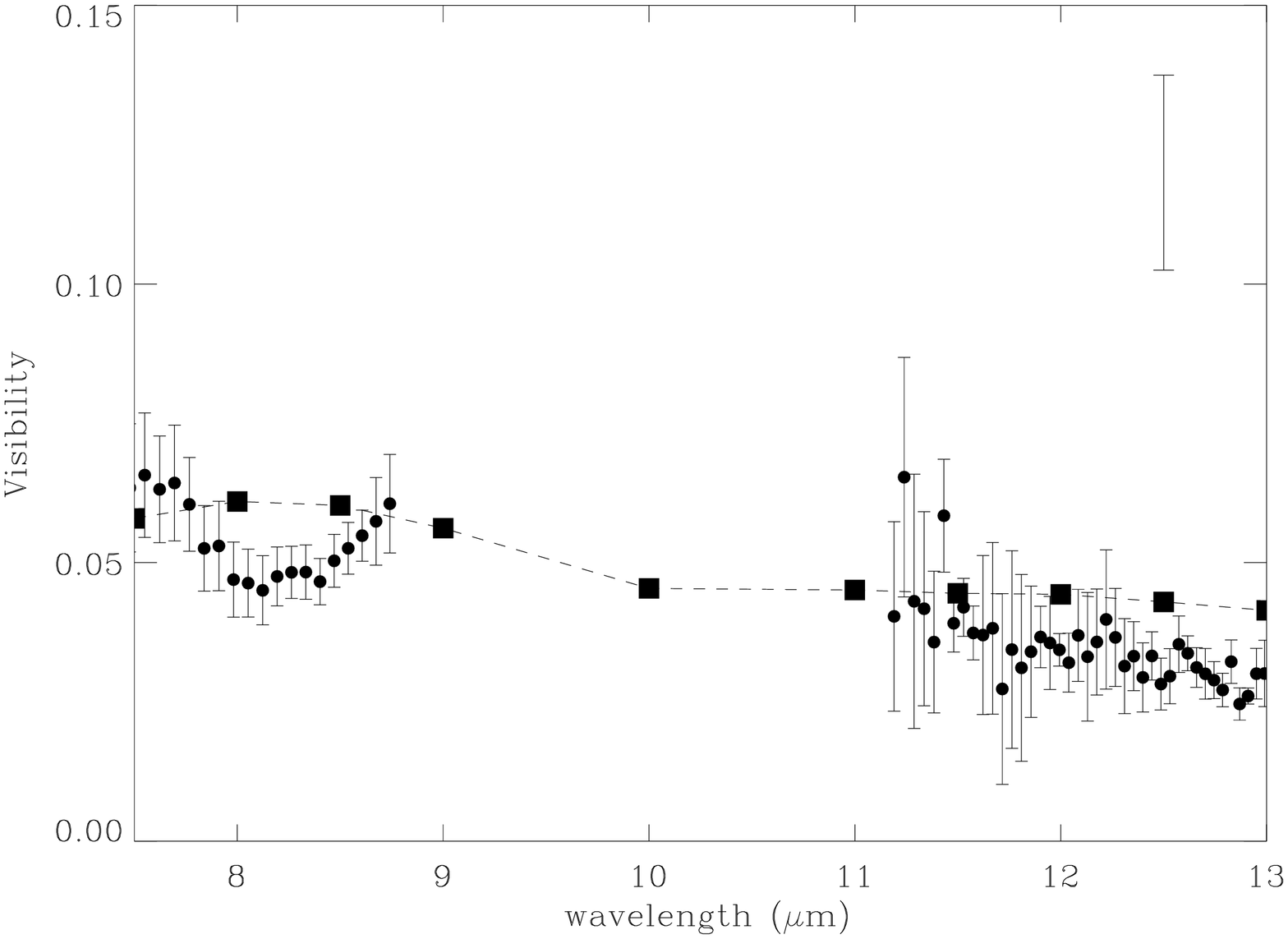}
  \includegraphics[height=9cm,width=5.5cm,angle=90]{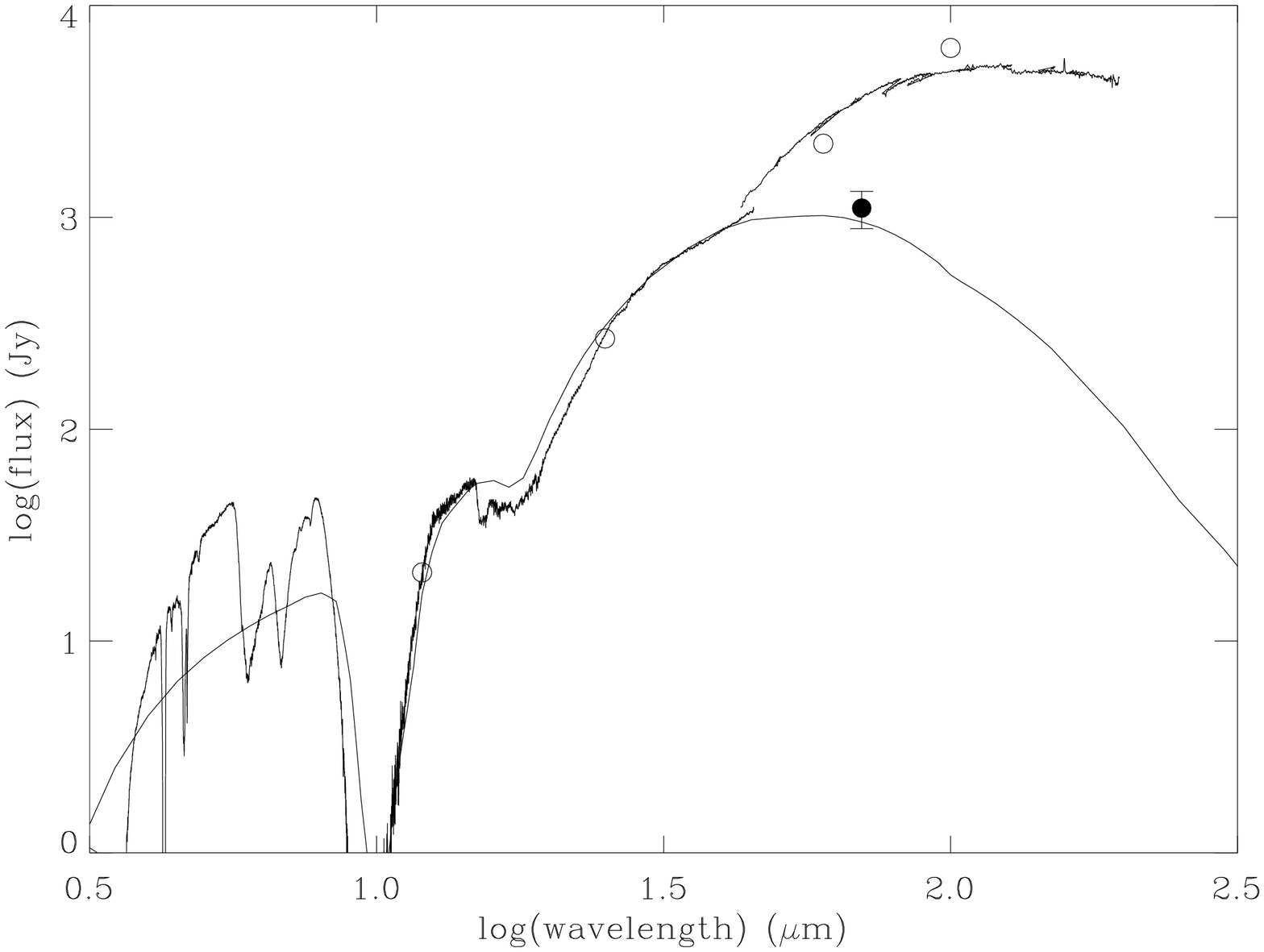}
  \caption[]{DUSTY model fits to visibilities and SED. Model is described by a $r^{-0.5}$
density law, {\bf and $A_{V}=100$}, star parameters are $L=4\, 10^4 L_{\odot}$,
$T_{eff}=10\,000$K, and a distance of 3800\,pc. {\it
Left:} Calibrated MIDI visibilities of W33A (dots with errorbars). The level
of systematic uncertainty is indicated by the bar in the right top
corner. Black squares connected by the dashed line correspond to the best-fit model. {\it Right:} SED of W33A traced by ISO-SWS and LWS spectra and
IRAS points (open circles). Full line is the model fit. IRAS/ISO large beam
effect at the longer wavelength is illustrated by the improved photometry
using MIPSGAL images (filled circle).}
  \label{Vis}
\end{figure*}

The corresponding visibility spectrum is obtained by dividing the
correlated flux by the flux spectrum and is presented in left panel of
Fig.\,\ref{Vis}.  Visibilities are not plotted when corresponding
to correlated fluxes smaller than 0.1Jy, a value below which the
measurements become unreliable (Jaffe et al. 2004, Matsuura et
al. 2006). The visibility spectrum shows that the silicate wings are
not strongly affected by the decrease in flux but instead follow the
declining trend of the continuum visibilities. If we would represent
the emission by a Gaussian emitting distribution then the FWHM size
increases from 30\,mas at 8$\mu$m to 60\,mas at 13$\mu$m.

\section{Radiative Transfer modelling}

Previous model fits of MYSO SEDs have indicated that simple spherical
radiative transfer models with roughly constant densities best fit the
near-IR and far-IR data (Churchwell et al. 1990; G\"{u}rtler et
al. 1991\nocite{1991A&A...252..801G}).  However, evidence exists that
the outer envelopes (10,000\,AU scales) have steeper density profiles
(Hoare et al. 1990\nocite{1990MNRAS.244..193H}; van der Tak et
al. 2000). We explore here whether the visibilities produced by
material on 100\,AU scales (a scale previously unexplored), and the
SED can be simultaneously matched by simple 1D spherically symmetric
dust models.  Arguably, such models are inadequate for MYSOs which are
likely to consist of circumstellar disks, bipolar cavities, etc.
Despite this, we explore whether the basic levels and trends of the
dispersed visibilities can be matched by a single unresolved star
deeply embedded in a dusty envelope before introducing more free
parameters.

For this purpose, we employ DUSTY, a code that solves the scaled 1D dust
radiative transfer problem (see Ivezic \& Elitzur
1997\nocite{1997MNRAS.287..799I}). We used a spherically symmetric
dust distribution illuminated by a central, unresolved star. The only
non-scaled parameter entering the code is the dust sublimation
temperature.  Gas emission is not taken into account by DUSTY.
Solutions are independent of the central source luminosity, and the
SED and visibilities can be scaled accordingly. The luminosity is the
prime stellar parameter that sets the inner dust sublimation radius
and thus the size scale; an increase causes the size of the emitting
region to be larger, as $r_{\rm sub} \propto \sqrt{L}$. The canonical
parameters for W33A are $L=1\,10^{5}\,L_{\odot}$, 3.8\,kpc
(Fa\'{u}ndez et al. 2004) and we initially chose $T_{\rm
eff}=25000$\,K typical for $15\,M_{\odot}$ star. These values make the
underlying star larger than a main sequence B1 star, consistent with
the idea that a star undergoing accretion at a high rate is swollen
(e.g. Behrend \& Maeder 2001\nocite{2001A&A...373..190B}). We tuned
the model output to the observations by varying DUSTY's input
parameters, {\it viz.} radial density distribution, dust properties,
and source $T_{\rm eff}$. The challenge is to construct a single model
that fits the observed visibilities, silicate wings, and mid-IR SED up
to 100$\mu$m.  We restrict the SED fitting to this wavelength
interval as it is well-known that 1D models have particular difficulty
in reproducing the short wavelength region because of the sensitivity
to the viewing angle (cf. Yorke \& Sonnhalter
2002\nocite{2002ApJ...569..846Y}).

We adopt a dust sublimation temperature of 1500\,K and an MRN-DL dust
mixture (Mathis et al. 1977\nocite{1977ApJ...217..425M}; Draine \& Lee
1984\nocite{1984ApJ...285...89D}) with a typical interstellar
graphite-silicate composition.  The outer bound of the model is
set at 1000 times the dust sublimation radius,  where the
temperature corresponds to the presumed ambient temperature of the
ISM, between 10 and 25\,K. The precise value for the outer radius does
not affect our conclusions here. 

Previous studies have shown that W33A is best described by an $A_{V}$
between 100 and 200 (Capps et al.  1978\nocite{1978ApJ...226..863C};
G\"{u}rtler et al.  1991).  A first result is that the MIDI
visibilities cannot be reproduced for any density distributions with
power exponents between $-2.0$ and $0.0$ for $A_{V}$ between 100 and
200 and $L=1\,10^{5}\,L_{\odot}$. The model visibilities at 10$\mu$m
are too small by an order of magnitude. Decreasing $A_{V}$ is not a
solution for normal dust, because of  W33A's exceptionally strong
silicate absorption feature.  Average model sizes at 10$\mu$m of $\sim
200$AU are reached only if the luminosity is reduced to about a
third. Such a reduction is justified when we consider 70$\mu$m MIPSGAL
data that we have obtained from the  Spitzer archive. These data
are taken at a vastly superior resolution compared to both IRAS and
ISO and reveal the presence of at least three point sources and strong
diffuse emission  within the IRAS beam. For W33A, we measure a
70$\mu$m flux of $1.1\,10^{3}$Jy (20\% uncertainty), as shown in
Fig.\,\ref{Vis}. This indicates that the true luminosity is
significantly less than deduced from IRAS data.

Even with the luminosity reduced to $4\,10^{4}\,L_{\odot}$, thermally
supported cores with $r^{-2}$ dependency (Larson
1969\nocite{1969MNRAS.145..271L}) are incompatible with the observed
 trend of decreasing visibilities with wavelength and the SED.
Incompatible models are also found if we adopt a constant infall
velocity type distribution with $r^{-1.5}$ (Shu
1977\nocite{1977ApJ...214..488S}). Reasonable fits to the SED that
produce sizes comparable to the MIDI visibilities are found for $r^{-1.0}$
logatropic distributions (e.g. McLaughlin \& Pudritz
1996\nocite{1996ApJ...469..194M}), but again they do not reproduce the
observed decreasing visibility trend with wavelength. DUSTY produces
smaller sizes, and thus larger visibilities, for longer
wavelengths. This affects especially the red wing of the silicate
feature, due to the diminishing opacity further out in the silicate
wing.
This mismatch in visibilities is however removed if we reduce the
slope to a much shallower density distribution with a $r^{-0.5}$ law
or a constant density law. These distributions fit the short
wavelength region  of the SED increasingly worse because of the
loss of warm dust. These shallower density distributions also require
the luminosity  of the central object to be fainter than
sB9observed.
We thus find that for standard ISM dust, the SED and the sizes at scales of
100\,AU limit the dust to follow distributions between $r^{-1.0}$ or
$r^{-0.5}$ powerlaws. The required optical depths are relatively high, yet the
best models do not fit the SED particularly well.

Different dust compositions influence both the magnitude and shape of
the visibility spectrum. We now explore dust made of cold and warm
silicates by Ossenkopf et al. (1992\nocite{1992A&A...261..567}) and
amorphous carbon. Ossenkopf silicates have the strong advantage of a
larger optical depth in the silicate feature than the DL grains. 
This property is advantageous in case of W33A, because it allows
models with relatively moderate $A_{V}$ to produce deep silicate
absorption. Models with moderate extinction fit the shorter wavelength
fluxes better. We find that the typical ISM ratio of 0.88
between graphite and silicates does not produce enough silicate
absorption to fit W33A's strong absorption feature. A better
correspondence is reached if this ratio is reduced to 0.50 (see also
Churchwell et al. 1990). The warm Ossenkopf silicates bring the extra
advantage of somewhat reducing the sublimation radius.
Although the reduction in $A_{V}$, and a better correspondence to the
silicate feature depth produces reasonable correspondence between the shallow
density models with the observation, models that fit the data better are found
by lowering the $T_{\rm eff}$ of the star. The change of this parameter produces
a decrease of the size of the inner dust boundary, because of the lower dust opacity with
temperature. We finally arrive at a simultaneous fit  (Fig.\,\ref{Vis}) to
the red silicate wing, SED peak and visibilities for a central
object with a relatively low $T_{\rm eff}$ of $10^{4}$\,K (corresponding
to a B9 supergiant, again consistent with the notion that an accreting star may be swollen) and a $r^{-0.5}$ density law (Fig.\,\ref{Vis}).  The
dust sublimation radius for $T_{d}=1500$\,K is found at 7 mas  (26.5 AU).

In summary, fitting 1D DUSTY models to the visibilities and SED
of W33A has shown that for nominal stellar parameters, size scales of
the emission region are too large and produce the wrong trend with
wavelength. Shallow radial density distributions produce this trend of
size with wavelength and in order to fit the depth of the silicate
feature as well, dust models with warm Ossenkopf silicates with an
increased silicate over graphite ratio reproduce the observation best,
provided the luminosity and $T_{\rm eff}$ of the central object are
reduced.

\section{Conclusions}

We presented mid-IR high-resolution dispersed interferometric
observations of the forming massive star W33A. The visibility spectrum
indicates an equivalent Gaussian FWHM of the emitting region of $\sim
120$AU, at 8$\mu$m, increasing to $\sim 240$AU, at 13$\mu$m. We
interpreted the interferometric data with simple spherically symmetric
DUSTY models  representative of a (unresolved) star embedded in a
dusty envelope, aiming for a simultaneous fit to the SED, silicate
profile and visibility spectrum. For any radial dust distribution we
found that the canonical value of W33A's luminosity is not compatible
with the visibilities. This is supported by MIPSGAL data. We found
that even for a reduced luminosity, the model produces too large
emitting regions causing the visibilities and SED to be mutually
incompatible. Changing the dust composition improves the situation for
an increased graphite to silicate ratio and using warm silicate
optical constants from Ossenkopf et al. (1992). We resorted to a
substantial lowering of the $T_{\rm eff}$ in order to obtain a
satisfactorily match between observables and models.  Further coverage
of the uv-plane will be very rewarding, constraining more appropriate,
higher dimensionality models, which will eventually lead in a proper
description of the circumstellar environment of an accreting massive
star. A full 2D axi-symmetric treatment of this and other MIDI data
will be the subject of a future paper.

\begin{acknowledgements} It is a pleasure to thank Olivier Chesneau and Rainer 
K\"{o}hler for discussion on the MIDI data reduction.
\end{acknowledgements}

\end{document}